\documentclass[prl,aps,twocolumn,groupedaddress,floats,final,nopacks,superscriptaddress]{revtex4-2}
\usepackage{graphicx}
\usepackage{subfigure}
\usepackage{amssymb}
\usepackage{latexsym}
\usepackage{epsfig}
\usepackage{psfrag}
\usepackage{dcolumn}
\usepackage{amsmath,amsfonts,amssymb,bm,ulem}
\usepackage{color}
\usepackage{float}
\usepackage[dvipsnames]{xcolor}
\usepackage[colorlinks=true, linkcolor=blue, citecolor=blue, urlcolor=blue]{hyperref}

\newcommand{\be}{\begin{equation}}
\newcommand{\ee}{\end{equation}}
\newcommand{\bea}{\begin{eqnarray}}
\newcommand{\eea}{\end{eqnarray}}

\begin{document}
\title{Avoided Stoner instability at a single ordinary Van Hove point}

\author{I.S. Tupitsyn}
\affiliation{Department of Physics, University of Massachusetts, Amherst, MA 01003, USA}
\author{B. Currie}
\affiliation{Physics Department, King's College London, Strand, London WC2R 2LS, UK}
\author{A.V. Chubukov}
\affiliation{W.I. Fine Theoretical Physics Institute and School of Physics and Astronomy, University of Minnesota,
Minneapolis, Minnesota 55455, USA}
\author{B.V. Svistunov}
\affiliation{Department of Physics, University of Massachusetts, Amherst, MA 01003, USA}
\author{E. Kozik}
\affiliation{Physics Department, King's College London, Strand, London WC2R 2LS, UK}
\author{N.V. Prokof'ev}
\affiliation{Department of Physics, University of Massachusetts, Amherst, MA 01003, USA}

%%%%%%%%%%%%%%%%%%%%%%%%%%%%%%%%%%%%%%%%%%%%%%%%%%%%%%%%%%%%%%%%%%%%%%%%%%%%%%%%%%

\begin{abstract}
When the Fermi surface and the Brillouin zone boundary touch at a Van Hove point, mean-field analysis predicts a ferromagnetic (Stoner) instability at finite $T_{MF}$ for any coupling strength due to the divergent density of states. However, the predicted effect has not been observed experimentally. Several qualitative theoretical proposals have been put forward to explain why the mean-field prediction fails. Based on numerically exact results for the two-dimensional Hubbard model with an ordinary Van Hove singularity, we uncover the mechanisms behind the suppression of the ferromagnetic instability. We employ two diagrammatic Monte Carlo approaches: (i) the four-channel self-consistent approximation and (ii) numerically exact method of combinatorial summation of diagrams with controlled resummation of the truncated expansion. We find that the system avoids the Stoner instability down to temperatures an order of magnitude below $T_{MF}$ due to the combination of the downward renormalization of the effective coupling and the suppression of the density of states by the loss of the quasiparticle residue.
\end{abstract}

\maketitle

%%%%%%%%%%%%%%%%%%%%%%%%%%%%%%%%%%%%%%%%%%%%%%%%%%%%%%%%%%%%%%%%%%%%%%%%%%%%%%%%%%

%%%%%%%%%%%%%%%%%%%%%%%%%%%%%%%%%%%%%%
\textit{Introduction---}Stoner ferromagnetism~\cite{Stoner1939} is perhaps the most studied, yet still not fully understood  instability of a Fermi liquid with repulsive interaction between fermions (see, e.g., ~\cite{Kanamori1963,Shimizu1964,Lifsic2006,Dzero2004,Moriya2012,Conduit2010}). Recent interest in this phenomenon~\cite{Bultinck2020,Vu2021,Chatterjee2022,  Blinov2023,*Blinov2023a,Antebi2023,Zach2024PRL,*Zach2024PRB,*Zach2024a,*megan,*Mayrhofer2025,Rakhmanov2023,Dong2023,*Dong2023a,*Dong2023aa,*Dong2023b,Szabo2022} is triggered by the experimental discoveries of spin and valley ferromagnetism in multi-layer graphene systems~\cite{Zhou2022,Holleis2025,DeLaBarrera2022}, transition metal dichalcogenides~\cite{Ghiotto2024}, and in quantum well AlAs~\cite{Hossain2020,*Hossain2021}.

A ferromagnetic transition can be either discontinuous (first order)  or continuous (second order). For the latter, a Stoner instability is indicated by a divergence of the static susceptibility at zero momentum, $\chi (T)$. Within mean-field approximation, $\chi^{-1} (T)$ in a  system with a repulsive on-site interaction $U$ scales as $1 + U \Pi_0 (T)$, where $\Pi_0 (T) <0$ is a static uniform polarization bubble of a Fermi gas. The instability towards ferromagnetism develops at $T_{MF}$ at which $U  \Pi_0 (T_{MF})  = -1$. When the Fermi surface (FS) and Brillouin zone (BZ) boundary touch at a single point, $\mathbf{Q}_{VH}$ [an ordinary Van Hove (VH) point],  $\Pi_0 (T)$ diverges as $\ln(T/E_0)$ with $E_0 \lesssim E_F$, where $E_F$ is the Fermi energy,
due to logarithmic divergence of the density of states (DOS), see, e.g., Refs.~\cite{Dzyaloshinskii1987,*Dzyaloshinskii1988,Wang2013, Betouras2020,*Efremov2019,Alvarez1998,*Honerkamp2001,*Hlubina1996,*LeHur2009,Chichinadze2020a,*Chichinadze2022,*Chichinadze2022a}.
By this logic, at a low enough $T$, the condition for Stoner ferromagnetism is  satisfied for any $U>0$ \cite {Irkhin2001,Zhang2021}.

However, this reasoning is not supported by quantum Monte Carlo (MC) simulations~\cite{Ma2010}, which did not observe ferromagnetism for an ordinary Van Hove singularity and only reported stronger ferromagnetic response at lower $T$, and with experiments on, e.g., Sr$_2$RuO$_4$ under uni-axial compressive strain, which found superconductivity instead of ferromagnetism at a Van Hove point~\cite{Hicks2014,Barber2018,*Stangier2022,*Li2022}. Theoretically, no Stoner instability has been detected at a Van Hove point in renormalization group (RG) studies \cite{RG1,RG2,Andrey2024}. The authors of Refs.~\cite{RG1,RG2,Andrey2024}
argued that the reason is a downward renormalization of $U$  from scattering in the particle-particle channel. However, a conclusive understanding of the mechanisms behind the avoided ferromagnetism is still lacking, while this physics must ultimately underpin the superconducting state in Sr$_2$RuO$_4$~\cite{Hicks2014,Barber2018,*Stangier2022,*Li2022}.

The beyond-mean-field scenario of Ref.~\cite{Andrey2024} consists of a renormalization of the coupling due to contributions from the crossed bare particle-hole bubble to the particle-hole susceptibility. Renormalization of $U$ by one such insertion contains the same single logarithm as $\Pi_0 (T)$ and transforms it into $U^* = U +  b U^2 \Pi_0(T)$ with $b\approx 1.88$. It was then conjectured that this result could be generalized to geometric series and proposed that the full renormalization of $U$ is
\begin{equation}
U^*/U = \left( 1 - U b\Pi_0(T) \right)^{-1} .
\label{mcrossed}
\end{equation}
Since $b > 1$, the Stoner condition  $U^* \Pi_0(T) =-1$ is not satisfied for any $U$, no matter how large. It is, however, \textit{a priori} unclear how to correctly sum up the logarithmic corrections in the particle-particle channels and whether it is even justified to restrict the renormalization of $U$ to maximally crossed diagrams. Furthermore, the relevance of the Fermi liquid description at the VH point was not questioned.

In this Letter,  we  address this problem within the controlled diagrammatic Monte Carlo framework~\cite{Prokofev2007, VanHoucke2010, Kozik2010,Kozik2024}. We find that the Stoner instability is avoided down to the lowest accessible temperature---about an order of magnitude below the mean-field critical temperature $T_{MF}$---and uncover the mechanisms responsible for the avoided ferromagnetism, which turn out to be more intricate than previously suggested. Following Refs.~\cite{RG1,RG2,Andrey2024}, we express the spin susceptibility as
\begin{equation}
\chi (T) = -\frac{\Pi (T) }{1+U^* (T) \Pi (T)} \; .
\label{Ueff}
\end{equation}
In contrast to the conventional definition of $\chi$ in a Fermi liquid, in which vertex corrections are included in the polarization bubble, we define $\Pi$ as a convolution of the two dressed fermionic propagators and absorb vertex corrections into the dressed $U^* (T)$ (see, e.g., ~\cite{Lifsic2006,Chubukov2018}). Our $U^*$ thus captures the net effect of the vertex corrections to the particle-hole bubble, which tend to increase it, and renormalizations from the particle-particle channel, which tend to decrease it -- treating both on equal footing. We find that the absence of the Stoner instability is driven by a combination of two effects. First, while $\Pi(T)$ still increases in magnitude as $T$ decreases, it is significantly reduced compared to $\Pi_0(T)$ due to the suppression of the quasiparticle residue. This residue becomes strongly frequency-dependent, indicating that the system develops non-Fermi-liquid behavior. Second, the downward renormalization of $U^*$ from the particle-particle channel outweighs the upward renormalization from vertex corrections, causing $U^*(T)$ to decrease overall. The net result is that $U^*|\Pi(T)|$ remains smaller than one down to the lowest temperatures in our simulations.

\begin{figure}[t]
\begin{center}
\includegraphics[scale=0.38]{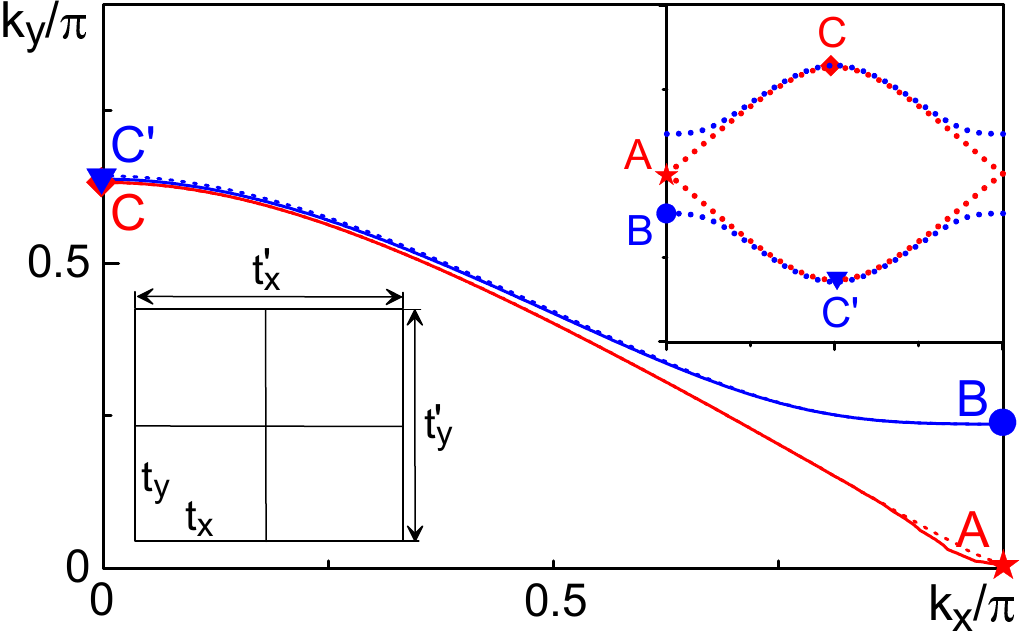}
\end{center}
\vspace{-5mm}
\caption{
Fermi surfaces of the interacting (solid lines) and non-interacting (dashed lines) systems for the same system densities. In the interacting case, at $\mu=\mu_{VH}$ the ${\mathbf Q}_{VH}$ point is located on the FS (point A); when $\mu$ is detuned from $\mu_{VH}$ by $0.2t$, we are dealing with an open FS intersecting BZ boundary at point B. Along $y$-direction the FS is located far from the BZ boundary (points C and C$'$ correspond to $\mu=\mu_{vH}$ and $\mu=\mu_{VH}+0.2t$ cases, respectively). Upper inset demonstrates full Fermi surfaces for both non-interacting cases. We show results calculated at $T/t=0.01$ when $\mu_{VH}(U,T) = -0.110 t$ (for non-interacting case $\mu_{VH}=0.76t$). Lower inset shows our hopping scheme.
}
\label{Fig1}
\end{figure}

%%%%%%%%%%%%%%%%%%%%%%%%%%%%%%%%%%%%%%
\textit{Model---}We consider an anisotropic Fermi-Hubbard model on a square lattice defined by
the Hamiltonian
\begin{equation}
H = - \sum_{i,j, \sigma} t_{i,j} \hat{c}^{\dag}_{i,\sigma} \hat{c}^{\,}_{j,\sigma}
+ U \sum_{i} \hat{n}_{i,\uparrow} \hat{n}_{i,\downarrow} - \mu \sum_{i,\sigma} \hat{n}_{i,\sigma},
\label{HHubb}
\end{equation}
where $\hat{c}^{\dag}_{i,\sigma}$ creates a fermion with spin projection $\sigma \in \{\uparrow,\downarrow\}$ on site $i$ and $\hat{n}_{i,\sigma}=\hat{c}^{\dag}_{i,\sigma} \hat{c}^{\,}_{i,\sigma}$ is the number operator. We chose the bare dispersion in the form $\epsilon_{\mathbf k} = - 2\sum_{\alpha=x,y} [t_{\alpha} \cos{k_{\alpha}} - t{'}_{\alpha} \cos{2k_{\alpha}}]$ with the nearest-neighbor, $t_x=t/2, \; t_y=t$, and next-nearest-neighbor, $t^{'}_x=0.12t, \; t^{'}_y=-0.24t$, hopping amplitudes (see lower inset in Fig.~\ref{Fig1}) designed to increase the density of states $\rho_F$ at the ordinary Van Hove point (below we take $t$ as the unit of energy). Graphically, this dispersion leads to an asymmetric FS, ``flattened" along the $y$-axis (see upper inset in Fig.~\ref{Fig1}). The chosen dispersion relation ensures that $T_{MF}$ remains reasonably high for relatively weak Hubbard repulsion set at $U=2t$ when other interaction-induced effects---such as competing instabilities and changes in the dispersion relation (and thus FS shape)---are small.
\begin{figure*}[t]
\centering
\includegraphics[width=0.325\linewidth]{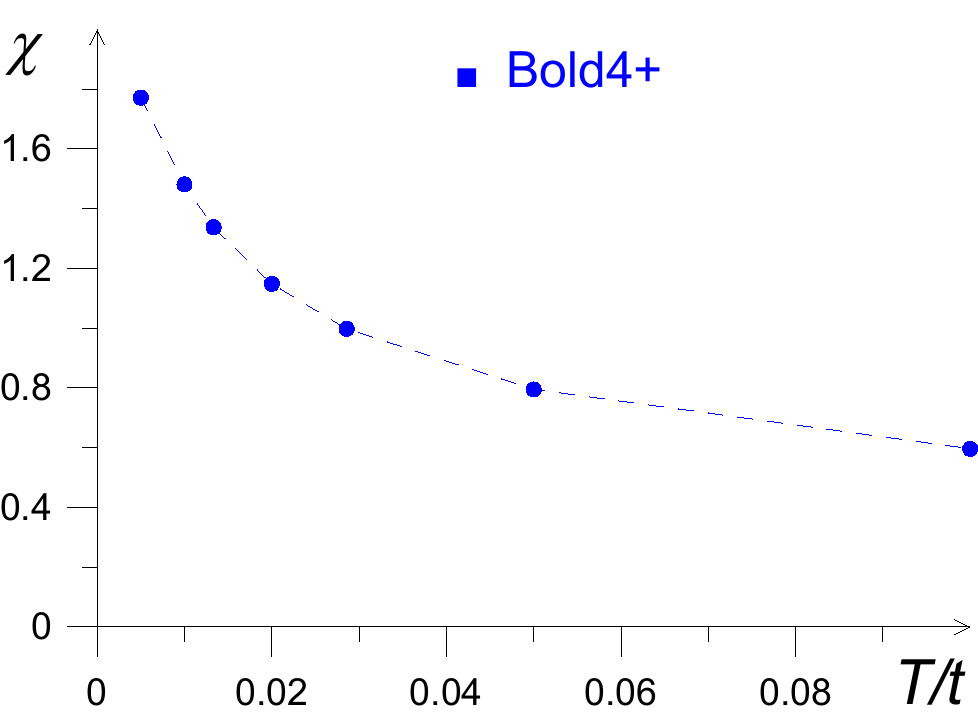}
\includegraphics[width=0.325\linewidth]{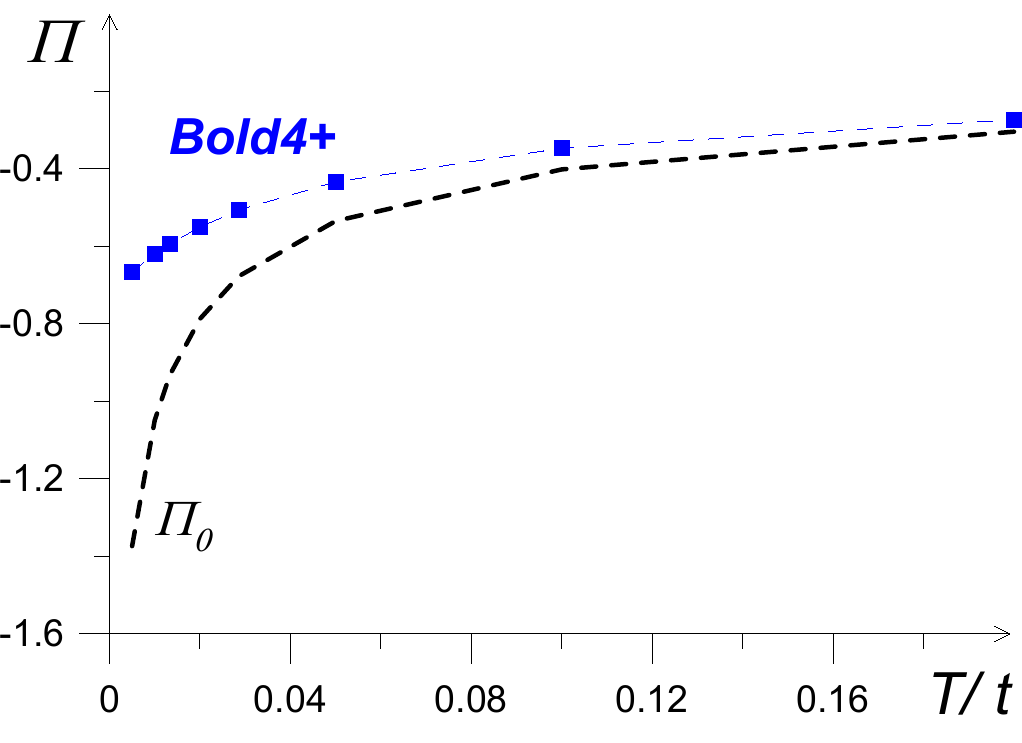}
\includegraphics[width=0.338\linewidth]{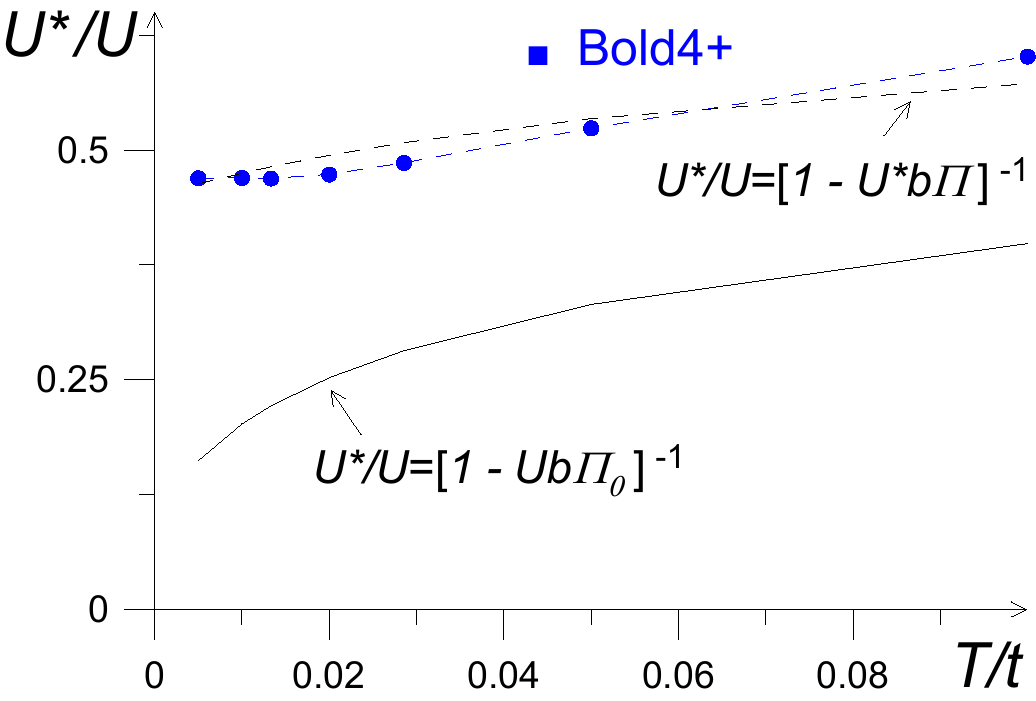}
\caption{
Bold4+ approximation results for $U=2t$ at Van Hove singularity. Left panel: static and uniform susceptibility.
Middle panel: static polarization bubble diagram at zero momentum computed using dressed Green's functions; in the low-temperature limit, it saturates to a finite value in contrast with the divergent behavior of $\Pi_0$ based on the bare
Green's functions (dashed line). Right panel: effective interaction extracted from the magnetic susceptibility
and $\Pi$ using Eq.~(\ref{Ueff}). The black dashed line shows $U^*/U$ corresponding to Eq.~(\ref{scU}).
}
\label{Fig2}
\end{figure*}

\begin{figure}[h]
\begin{center}
\subfigure{\includegraphics[scale=0.245]{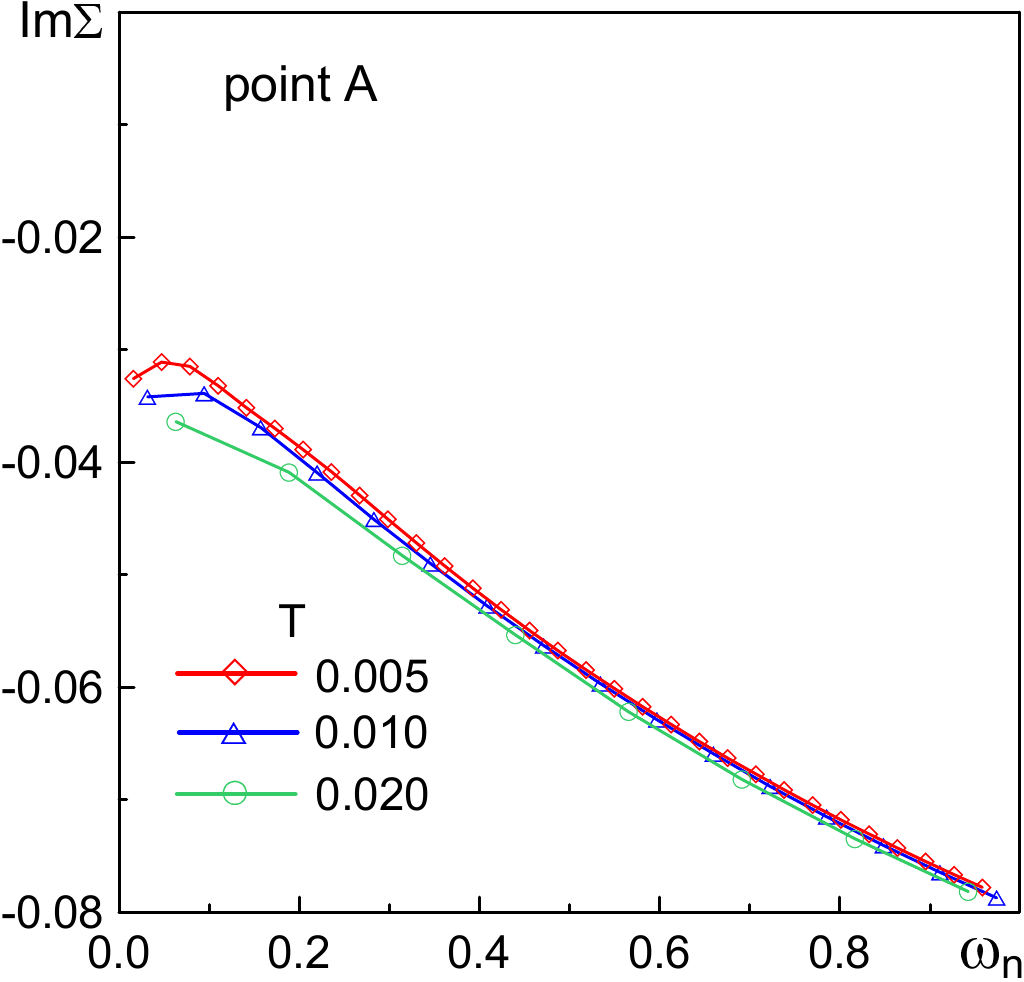}}
\subfigure{\includegraphics[scale=0.245]{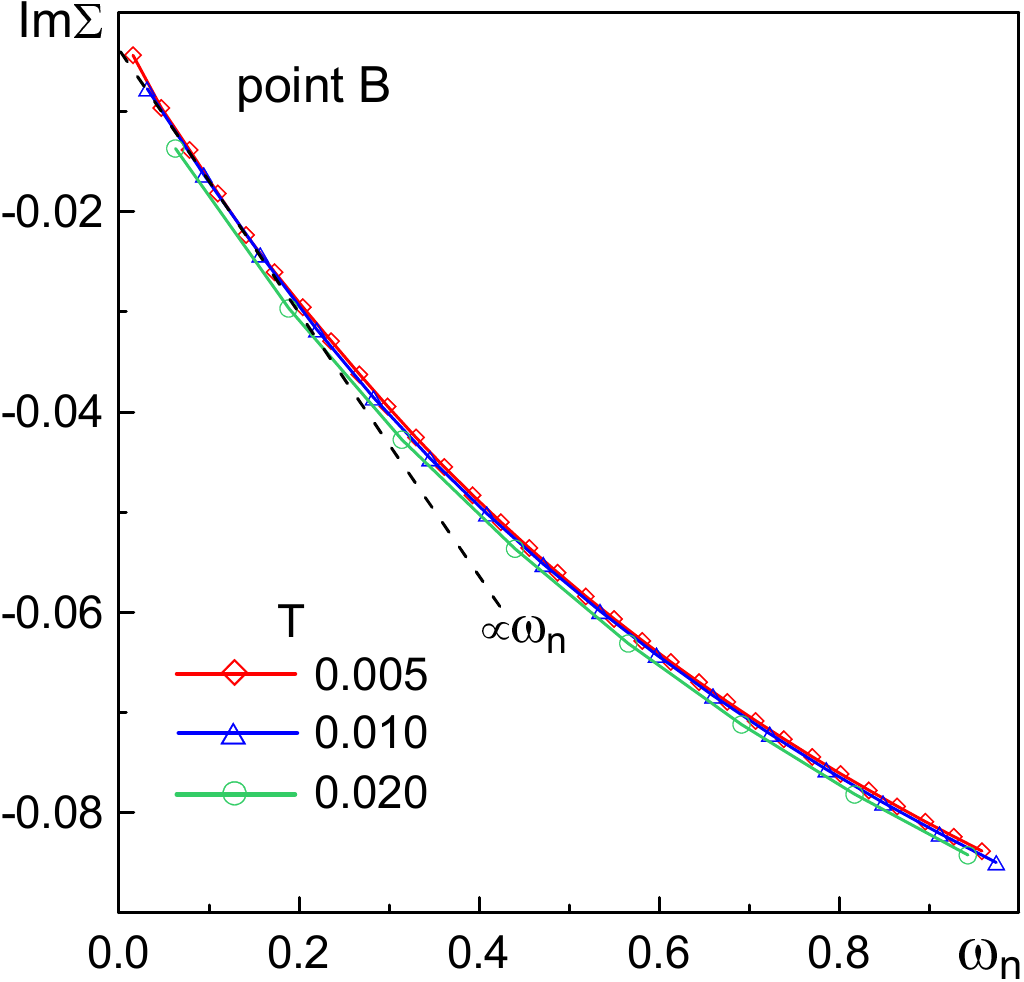}}
\end{center}
\vspace{-5mm}
\caption{Bold4+ results for the frequency dependence of $\Sigma ({\mathbf Q}_{VH},\omega_n)$ at different temperatures for the cases  when the FS touches the BZ boundary at the point A in Fig.~\ref{Fig1} (left panel),
and when the  system is away from Van Hove point and the system has an  open FS, which intersects BZ at the point B in Fig.~\ref{Fig1} (right panel).
Error-bars are within the symbol sizes. The dashed line in the right panel is the
expected FL behavior $\Sigma (\omega_n) = -i \alpha \omega_n$
}
\label{Fig3}
\end{figure}

We compute the static susceptibility as $\chi (\mathbf{Q})= 2 \int d\tau \sum_{j} \exp(i\mathbf{r}_j \mathbf{Q}) \langle S_z(0, 0) S_z(\tau, \mathbf{r}_j) \rangle $ at $Q=0$. In Fig.~\ref{Fig1}, we show the actual FS for $U=2t$ and the one for free fermions, both at the Van Hove point and away from it. The FS is defined by the standard condition ${\rm Re} \, G^{-1} (\omega \to 0, {\mathbf k}) = 0$; the quasiparticle residue is defined as  $Z({\mathbf k}) =  \lim_{T\to 0}[1-{\rm Im} \Sigma(\pi T, {\mathbf k}) / (\pi T)]^{-1}$. We see that the interaction-induced change of the Fermi surface is negligibly small and can be safely neglected even at the Van Hove point, when  the FS and BZ touch at momentum ${\mathbf Q}_{VH}=(\pi/a, 0)$. We also verified (see Fig.~\ref{Fig3} right panel) that away from the FS, the system remains in a perturbative FL regime. At $\mu=\mu_{VH}$, a non-perturbative regime emerges at $T \lesssim T_{MF} \approx 0.065t$.

%%%%%%%%%%%%%%%%%%%%%%%%%%%%%%%%%%%%%%

We apply two diagrammatic MC (DiagMC) computational procedures. One, labelled Bold4+, is based on a fully self-consistent one-loop renormalization of the single-particle Green's function, $G$, and four-point vertices in the three two-body channels, and extended beyond one-loop to include the leading vertex corrections (for details see Fig.~6 in Ref.~\cite{MagCor2017} and the  End Matter section in Ref.~\cite{OurLieb}). It accounts for all diagrams up to fourth order in $U$ for self-energies, and some, but not all, higher-order diagrams in the form of embedded geometric series. This scheme is computationally efficient and allows us to explore what happens when logarithmic corrections from all channels are combined in a particular, Bold4+ scheme dictated, way.

The second approach employs numerically exact DiagMC framework~\cite{Prokofev2007, VanHoucke2010, Kozik2010}
to evaluate key properties (magnetic susceptibility $\chi$ and fermionic self-energy $\Sigma$) from their Taylor series expansions in $U$, $\sum_{n=0}^N  a_n U^n $, directly in the thermodynamic limit. We use the recently developed combinatorial summation (CoS) algorithm \cite{Kozik2024} to compute $a_n$ as sums of all connected Feynman diagrams of order $n$. The only sources of controlled errors are statistical uncertainties in $a_n$ from the MC integration over internal variables and the truncation of the Taylor expansion at a sufficiently large finite order $N$. The final answer is reconstructed using standard Dlog Pad\'e~\cite{Pade1961} and Integral Approximant methods \cite{Hunter1979IA, PadeSK2019}, which construct a sequence of approximants that converge at a given $U$. The spread among different approximants provides a reliable estimate of the systematic extrapolation error, as described in Ref.~\cite{PadeSK2019}. In this work, $N=8$ is sufficient to achieve the required accuracy.

The net result of computations is that there is no Stoner ferromagnetism down to the lowest accessible temperature, which is an order of magnitude smaller than $T_{MF}$. However, the evolutions of $U^*$  and $\Pi$ with temperature revealed by CoS are qualitatively different from those predicted by Bold4+. In Bold4+, both $\Pi (T)$ and $U^* (T)$ tend to finite values at $T \to 0$;  in CoS, $|\Pi (T)|$ keeps increasing and $U^* (T)$ keeps decreasing down to the smallest  $T$ in our simulations.  In both cases, the product $U^*|\Pi (T)| <1$. Perhaps the most unexpected result is prediction of the non-Fermi-liquid behavior of self-energy at VH point. Below we discuss the results of Bold4+ and CoS calculations separately.

\textit{Bold4+ results---}In the left panel of Fig.~\ref{Fig2}, we show $\chi (T)$ at $\mu=\mu_{VH}$ obtained within the Bold4+ scheme. We clearly see that $\chi (T)$ increases with decreasing $T$, but does not diverge down to the smallest $T$ in the simulations. In the middle and right panels of Fig.~\ref{Fig2}, we plot separately $\Pi (T)$ and the ratio $U^* (T)/U$ extracted using Eq.~(\ref{Ueff}). We see that $\Pi(T)$ does not show the divergent behavior of $\Pi_0 (T)$  and instead saturates to a finite value when $T \to 0$. We trace this behavior to the reduction of the fermionic residue. In Fig.~\ref{Fig3}, we show the fermionic self-energy $\Sigma (k,\omega_n)$ along the Matsubara axis for $\mathbf{k}$-points at the intersection between the FS and BZ labelled as points A and B in Fig.~\ref{Fig1}. Comparing the left (point A) and right (point B) panels, we see the dramatic effect of the Van Hove singularity: the FL behavior displayed by $\Sigma (\mathbf{k}_B, \omega_n)$  at low frequencies is replaced by $\Sigma (\mathbf{Q}_{VH}, \omega_n)$ saturating to a finite value at $\omega_n \to 0$.

\begin{figure*}[htbp]
\centering
\includegraphics[width=0.325\linewidth]{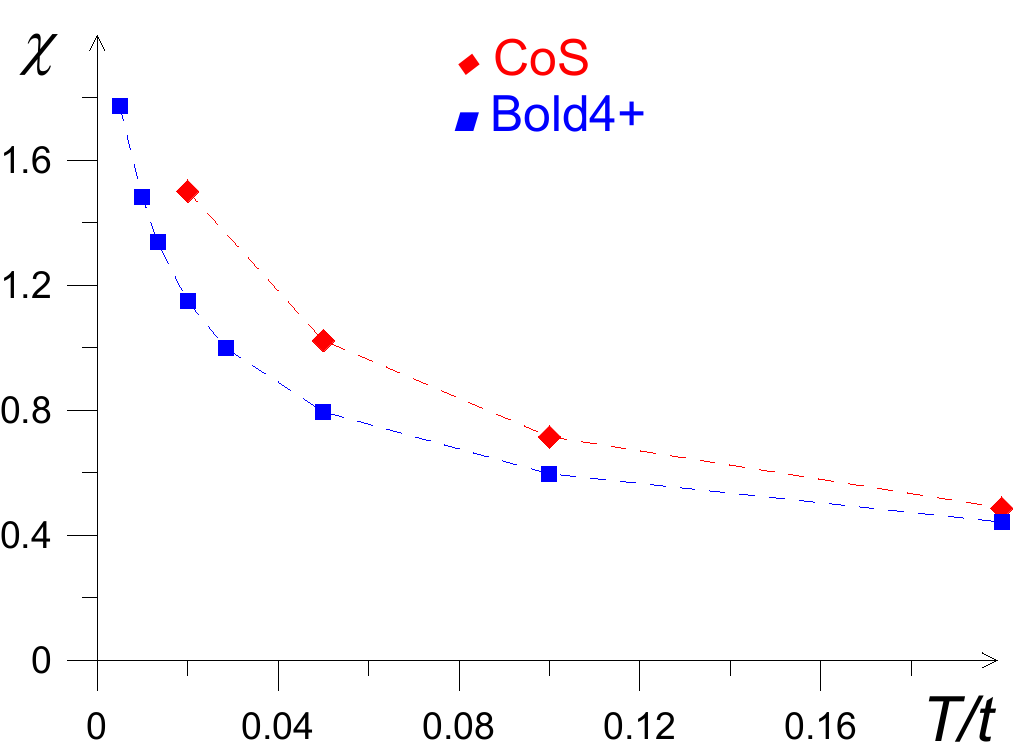}
\includegraphics[width=0.325\linewidth]{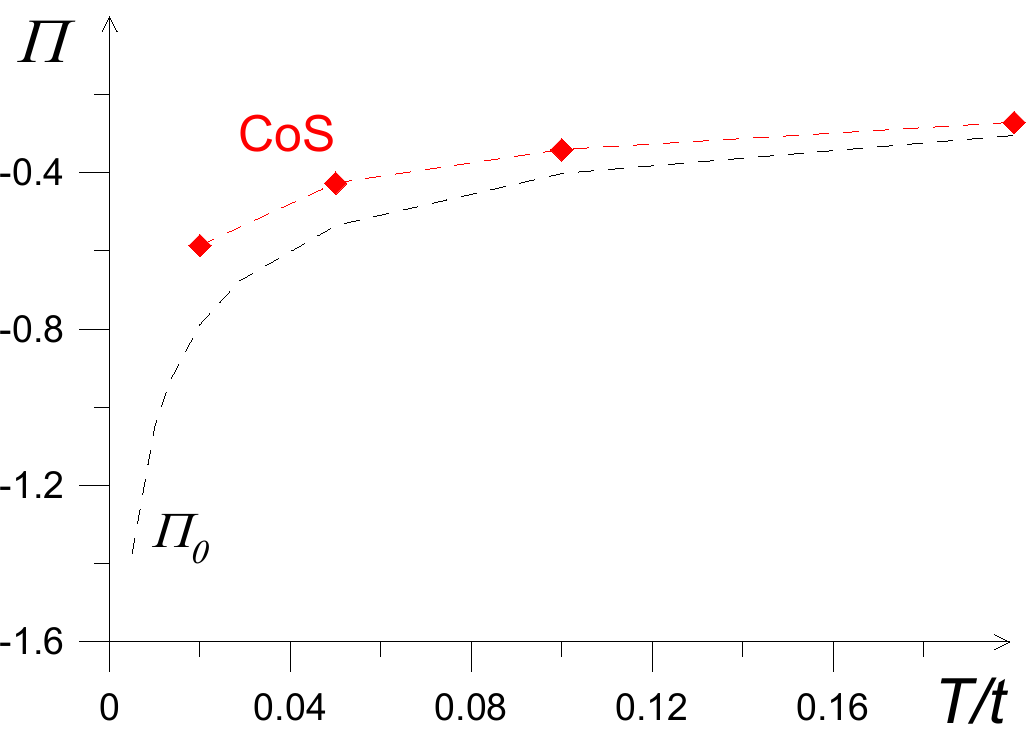}
\includegraphics[width=0.335\linewidth]{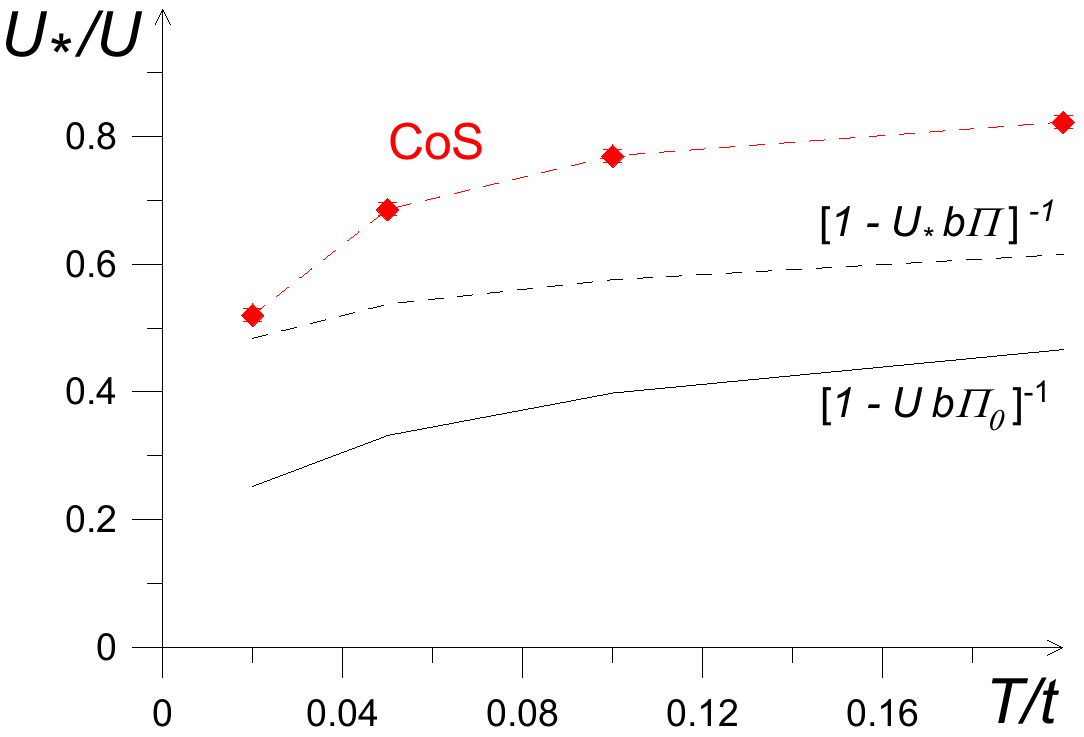}
\caption{
DiagMC-CoS results for $U=2t$ at Van Hove singularity. Left panel: static and uniform susceptibility (Bold4+ results from Fig.~\ref{Fig2} are added for direct comparison). Middle panel: static polarization bubble diagram at zero momentum computed using dressed Green's functions. It appears to diverge in the low-temperature limit though less dramatically when compared to the of $\Pi_0$ behavior shown by the dashed line. Right panel: effective interaction extracted from the magnetic susceptibility and $\Pi$ using Eq.~(\ref{Ueff}). By solid and dashed lines we plot predictions for $U^*/U$ in Eqs.~(\ref{mcrossed}) and Eq.~(\ref{scU}).
}
\label{Fig4}
\end{figure*}

In the right panel of Fig.~\ref{Fig2}, we show the result of the Bold4+ approach for $U^* (T)/U$ at the Van Hove point.
Curiously, this effective interaction can be reasonably well approximated by the solution of the self-consistent equation
\begin{equation}
U^*/U = \left[ 1 - U^* b  \Pi(T) \right]^{-1} \,,
\label{scU}
\end{equation}
which can be viewed as an {\it ad hoc} extension of  Eq.~(\ref{mcrossed}), where one includes the renormalization of the side vertices in the particle-hole bubble and computes it with the full Green's functions. The solution of (\ref{scU}) yields 
$U^*/U \sim (\Pi (T))^{1/2}$, weaker than proposed in~\cite{Andrey2024}. The ratio $U^*/U$ would still vanish at $T=0$  if $\Pi(T)$ was divergent. However, as we see from Fig.~\ref{Fig4}, at the lowest $T$, $\Pi (T)$ saturates at a finite value, hence $U^*/U$ also saturates.

%%%%%%%%%%%%%%%%%%%%%%%%%%%%%%%%%%%%%%
\textit{Controlled results---}In the left panel of Fig.~\ref{Fig4}, we present $\chi (T)$ obtained within the DiagMC-CoS scheme.  We again see that $\chi (T)$ increases with decreasing $T$, and does not diverge down to the smallest $T$ in the simulations.
In the middle and right panels of Fig.~\ref{Fig4} we present separately the values of the polarization bubble $\Pi (T)$ and the ratio $U^* (T)/U$ extracted from Eq.~(\ref{Ueff}).
\begin{figure}[thh]
\begin{center}
\includegraphics[scale=0.34]{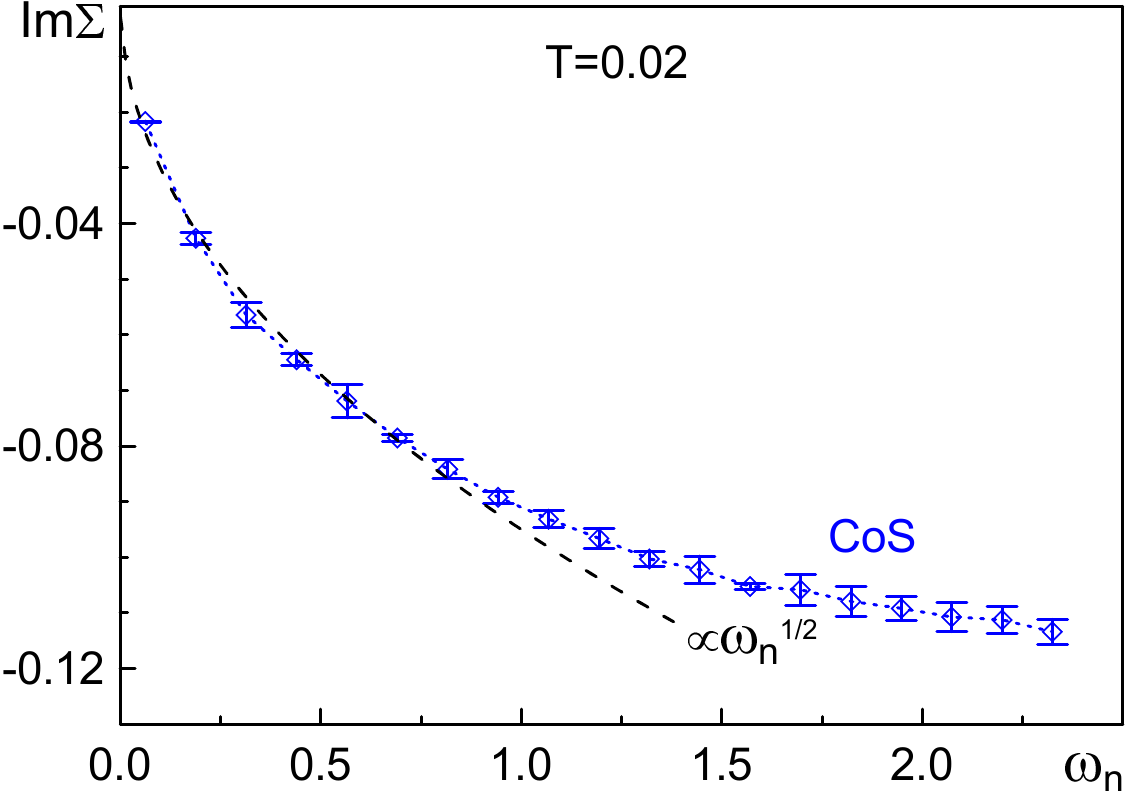}
\end{center}
\vspace{-5mm}
\caption{ DiagMC-CoS results for frequency dependence of $\Sigma ({\mathbf Q}_{VH},\omega_n)$, at $T/t=0.02$. The dashed line is a fit to non-FL $\omega_n^{1/2}$ dependence at small frequencies. }
\label{Fig5}
\end{figure}

We see that CoS results for $\Pi (T)$ and $U^*/U$ are different from Bold4+. That is, the static polarization bubble, shown in the middle panel of Fig.~\ref{Fig4}, continues to increase in magnitude as $T$ decreases, although the increase is weaker than that of  $\Pi_0$. We again can trace this behavior to properties of the fermionic self-energy $\Sigma ({\mathbf Q}_{VH},\omega_n)$, which at the Van Hove point displays a non-FL behavior  $\Sigma ({\mathbf Q}_{VH},\omega_n) \propto (\omega_n)^{\gamma}$ with $\gamma \approx 1/2$,  see Fig. \ref{Fig5}. At the same time, $U^*/U$, shown in the right panel of Fig.~\ref{Fig4}, decreases with lowering $T$ and does not saturate, in contrast to the results of the Bold4+ scheme. However, the renormalization of $U^*/U$ does not follow neither the logarithmic behavior in  Eq. (\ref{mcrossed}), nor the one in Eq. (\ref{scU}). Since our CoS data for ${\rm Im} \Sigma ({\mathbf Q}_{VH},\omega_n)$ show a non-FL power-law dependence at low frequencies---consistent with $\omega_n^{1/2}$, see Fig.~\ref{Fig5}---it is unlikely that $|\Pi (T)|$ will saturate at the lower temperatures beyond the reach of our simulations.

\begin{figure}[t]
\begin{center}
\includegraphics[scale=0.32]{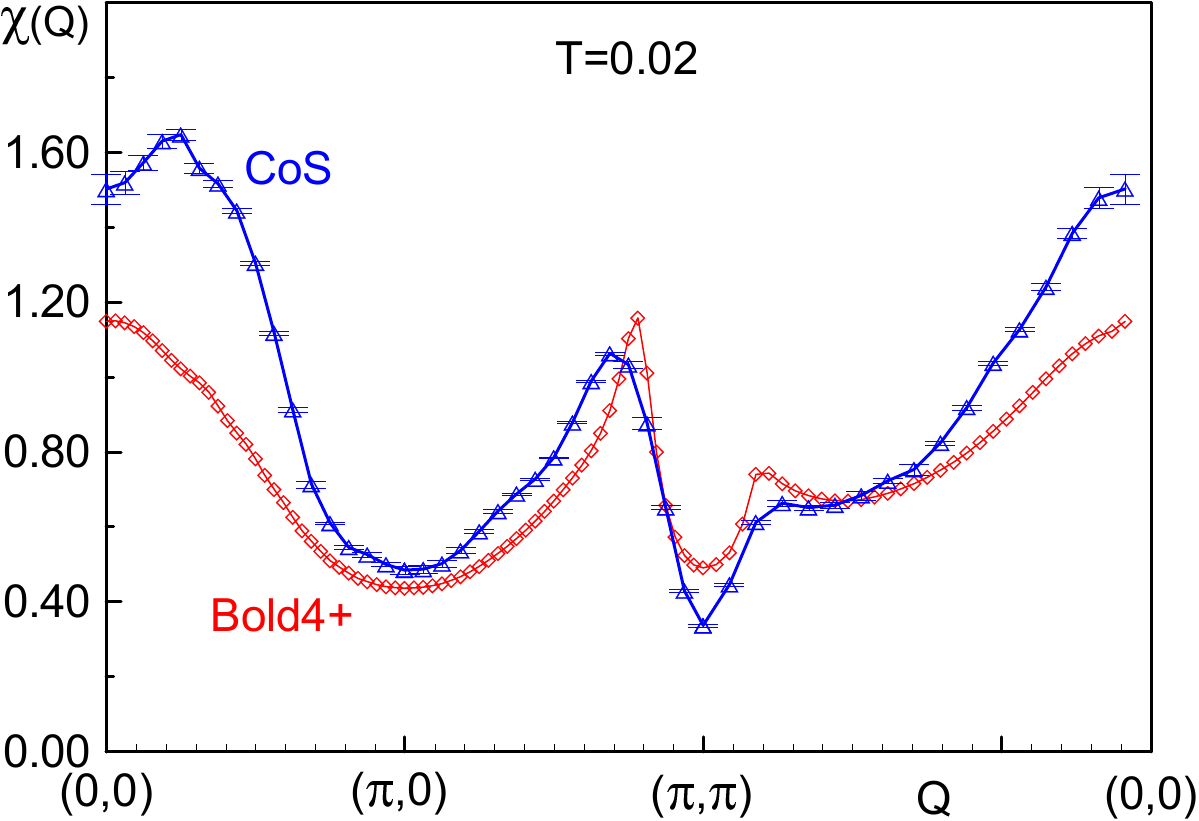}
\end{center}
\vspace{-5mm}
\caption{Static magnetic susceptibility $\chi(\mathbf{Q})$ along the $(0,0) \to (\pi ,0) \to (\pi,\pi) \to (0,0)$ path in BZ obtained within the CoS and Bold4+ techniques for $U=2$ and $T=0.02$ (for Bold4+ data, statistical error bars are within the symbol sizes). }
\label{Fig6}
\end{figure}

Finally, in Fig.~\ref{Fig6}, we present the momentum dependence of the static magnetic susceptibility $\chi (\bf {Q})$ at $T/t=0.02 \ll T_{MF}$ obtained in Bold4+ and CoS. Both techniques predict a modest enhancement of $\chi ({\mathbf Q})$ near  $\mathbf{Q}=0$, indicating that there are only short-range ferromagnetic correlations. An unexpected feature is the shift of the peak in $\chi ({\mathbf Q})$ from ${\mathbf Q}=0$ to finite $\mathbf{Q}$ at this temperature (at higher temperatures, the peak is at $Q=0$).   This may be the indication of  the non-analyticities in the momentum dependence of the static spin susceptibility, which were detected and analyzed in a 2D system away from a Van Hove point~\cite{Chubukov2009,Efremov2008}.
%

%%%%%%%%%%%%%%%%%%%%%%%%%%%%%%%%
\textit{Conclusions and outlook---} We addressed an issue crucial to the understanding of $\text{Sr}_2\text{RuO}_4$ and other materials: whether a two-dimensional (2D) Fermi liquid, tuned to a Van Hove singularity along a particular momentum direction, inevitably develops ferromagnetic order. Mean-field analysis predicts that this should be the case for any coupling strength due to the divergent density of states. However, evidence from experiments and quantum Monte Carlo simulations has called this understanding into question, while earlier beyond-mean-field considerations have remained inconclusive. We investigated the fate of the ferromagnetic instability using two distinct diagrammatic Monte Carlo approaches: Bold4+ and DiagMC-CoS. The former demonstrates that an advanced treatment of higher-order diagrams dramatically alters the physical picture compared to second-order perturbation theory, establishing that the problem is genuinely non-perturbative and cannot be captured solely by renormalizing the one- and two-body channels. To overcome this limitation, we deployed the DiagMC-CoS technique to sum all diagrammatic contributions up to high orders with controlled precision. The outcome of our analysis is that the reduction of the effective coupling, combined with the partial compensation of the divergent density of states by a reduced quasiparticle residue, suppresses the Stoner instability down to the lowest temperatures in our simulations. This reveals a physical scenario that contrasts with earlier suggestions~\cite{Andrey2024}.

A natural competitor to Stoner ferromagnetism at a VH singularity is superconductivity. In $\text{Sr}_2\text{RuO}_4$ tuned to a single VH point, superconductivity rather than ferromagnetism has been observed experimentally~\cite{Hicks2014,Barber2018,*Stangier2022,*Li2022}. Our insights into the mechanisms driving this avoided ferromagnetism can shed light on the nature of the resulting superconducting state. Furthermore, our approach could be extended to address the problem of ferromagnetic instability at an extended Van Hove point (a higher-order Van Hove singularity), where the density of states diverges as a power law~\cite{RG2}. This situation is notably realized in twisted $\text{WSe}_2$~\cite{guerci2024,christos2025,zang2021,devakul2021,varma2025}, where superconductivity rather than Stoner magnetism has likewise been observed~\cite{Mak_supercond_2024,Pasupathy2024superconductivity}.

\smallskip

\noindent \textbf{Acknowledgments.}
B.S. and N.P. acknowledge support from the National Science Foundation under Grant No. DMR-2335904. A.V.C. acknowledges support from the National Science Foundation grant NSF: DMR-2325357. I.S.T. and A.V.C. acknowledge support from Simons Foundation grant SFI-MPS-NFS-00006741-07 for the Simons Collaboration on New Frontiers in Superconductivity. B.C., E.K., N.P. and B.S. acknowledge support by the UK Engineering and Physical Sciences Research Council through grant EP/X01245X/1. The DiagMC calculations were performed using King's Computational Research, Engineering, and Technology Environment (CREATE).

\bibliography{refs.bib}

\end{document}